\documentclass[aps,prl,showpacs,amsmath,amssymb,twocolumn]{revtex4-1}
\usepackage{epsfig}
\usepackage{graphicx,color}
\usepackage{amsmath}
\newcommand{\beq}{\begin{eqnarray}}
\newcommand{\eeq}{\end{eqnarray}}

\begin{document}

\title{Two-step nuclear reactions: The Surrogate Method, the Trojan Horse Method and their common foundations}
\author{Mahir S. Hussein}
\affiliation{Instituto Tecnol\'{o}gico de Aeron\'{a}utica, DCTA,12.228-900 S\~{a}o Jos\'{e} dos Campos, SP, Brazil}
\affiliation{Instituto de Estudos Avan\c{c}ados, Universidade de S\~{a}o Paulo C. P.
72012, 05508-970 S\~{a}o Paulo-SP, Brazil}
\affiliation{Instituto de F\'{\i}sica,
Universidade de S\~{a}o Paulo, C. P. 66318, 05314-970 S\~{a}o Paulo, SP, Brazil}


\begin{abstract}
In this Letter I argue that the Surrogate Method, used to extract the fast neutron capture cross section on actinide target nuclei, which has important practical application for the next generation of breeder reactors, and the Trojan Horse Method employed to extract reactions of importance to nuclear astrophysics, have a common foundation, the Inclusive Non-Elastic Breakup (INEB)Theory. Whereas the Surrogate Method relies on the premise that the extracted neutron cross section in a (d,p) reaction is predominantly a compound nucleus one, the Trojan Horse Method, assumes a predominantly direct process for the secondary reaction induced by the surrogate fragment. In general, both methods contain both direct and compound contributions, and I show how theses seemingly distinct methods are in fact the same but at different energies and different kinematic regions. The unifying theory is the rather well developed INEB theory.
\end{abstract}

\maketitle

The recent upsurge of interest in two-step nuclear reactions stems from two reasons; application to reactor technology, and nuclear astrophysics. In the first, the Surrogate Method (SM), was proposed to eventually yield fast neutron compound nucleus cross sections of  its reaction with $^{238}$U, $^{232}$Th and other targets in the actinide region \cite{SM0, SM1, SM2}, of potential important practical use in future fast breeder reactors. The major motivation of the SM is the uncertainty about information obtained from a Hauser-Feshbach calculation of the CN reaction of interest owing to the uncertainty concerning the final decay channel's characteristics, density of states etc. To obtain the CN cross section, the SM actually measures the relevant CN reaction cross section of interest induced by the surrogate neutron in a $(d,p)$ reaction. This reaction supplies the "fast" neutrons while the proton is detected and its spectrum is analyzed and used to extract, say,  $n+ A\rightarrow CN \rightarrow y + B$. Thus the proton is detected in coincidence with $y$, or if the final channel is the fission of the CN, with one of the fission fragments. 

Of course the SM can be used in other reactions besides the (d,p). Let us call the projectile nucleus $a$, and the target $A$. Here $a$ is assumed to be composed of two fragments $a = b + x$. The detected fragment is $b$, while the fragment that interacts with $A$ and forms the compound nucleus is denoted by $x$.
The assumed form of the cross section for the case where the final decay channel of the compound nucleus, $(x + A)$ is $ y + B$ is $\sigma^{SM}_{(a, b)} = K_{SM}\times \sum_{l}[F_{(x +A, l)} \cdot G_{(f,l)}$, where $K_{SM}$ is a kinematical factor, $F_{(x + A, l)}$ is the formation probability of the compound nucleus $(x + A)$ at angular momentum $l$, while $G_{(i, l)}$ is the branching ratio for the decay of this compound nucleus into the final channel, $f$, at angular momentum $l$. Within the Hauser-Feshbach theory this branching ratio is given by $G_{(f, l)} = T_{(f, l)}/[\sum_{g} T_{(g, l)}]$, with $T$ indicating the transmission coefficient. The total compound nucleus cross section (fusion) is the inclusive sum over all final decay channel, $\sigma^{SM}_{capture} = K_{(SM)}\sum_{l} F_{(x + A, l)}$.  Going back to the $(d, p)$ reaction, it would be particularly useful to use the SM to obtain the capture cross section, $a + A \rightarrow CN \rightarrow (A + 1) + \gamma$. Unfortunately this particular CN decay  channel, the $\gamma$ emission leaving the (A+n) final nucleus in the ground state, is very sensitive to the spin distribution \cite{THM4}, making a SM determination of the capture cross section difficult. If the neutron capture cross section measurements were to be attained within the SM in the future it would be of great value in nuclear astrophysics as it supplies $n$ capture cross section important in element production through the s-process of supernova explosion. 

The second application is in the field of nucleosynthesis in nova evolution and the production of elements such as Sodium. Since the cross sections of reactions which yield these intermediate-light nuclei at the astrophysical energies of interest, of a few keV's, is very small, one relies on the so-called Trojan Horse Method (THM) \cite{Baur2003, THM1, THM2, THM3}. Within this method involving the reaction of a projectile, $a= b +x$, with a target, $A$, one is interested in, say, the direct reaction $x+ A \rightarrow y + B$, so the cross section is written as $\sigma_{THM}^{(b)} = K_{THM} \times |\phi(\textbf{k}_{b})|^{2}\times \sigma(x + A \rightarrow y +B)$, where $K$ is a kinematic factor, and $\phi$ is the internal wave function of the primary projectile, $a$, and $\textbf{k}_b$ is the momentum of the detected spectator fragment, $b$. The merit of the THM resides  the premise that since $x$ is brought about at the target position by the surrogate ion, $a$, most of the hindering effect of the Coulomb barrier is gone and the reaction $x + A \rightarrow y +B$ would proceed effectively above the Coulomb barrier. The other problem that complicates the measurement of, say, the reaction $x + A \rightarrow y + B$ at low energies for use in nuclear astrophysics is electron screening present  if $x$ is a primary projectile. However the THM supplies a secondary $x$ projectile at above barrier energies, as explained above, and accordingly the electron screening problem is avoided. These conditions, no Coulomb barrier to surpass, and no electron screening,  allows the extraction of the desired cross section, $x + A \rightarrow y + B$ through the reaction $a + A \rightarrow b + x + A \rightarrow b + y + B$, with relative ease even at the extremely low energies required to simulate the nova environment, $E_{a} = 100 - 400 keV$. As an examples, the reaction $^{18}F (p, \alpha)^{15}$O is the desired reaction and the measured one is $d(^{18}F, \alpha ^{15}$O)n, in the quasi elastic regime (higher energies and appropriate kinematics). The THM has been very useful in supplying astrophysical S-factor of relevant reactions in different scenarios and energies at which the direct measurements are either not feasible or do not exist.

I will argue in this note that the two seemingly different methods, SM and THM, have a common foundation; the Inclusive Non-Elastic Breakup theory (INEB). According to this theory \cite{UT1981, IAV1985, HM1985, Austern1987}, the cross section for the observation of the spectator fragment, $b$, is generally given by $\sigma^{(b)}_{(INEB)} = K_{(INEB)}\cdot \tilde{\sigma}^{(x + A)}_{R}$. The kinematical factor $K_{(INEB)}$ contains the density of states of the observed fragment, $\rho_{b}(E_b) = \mu_{b}k_{b}/[(2\pi)^{3}\hbar^3]$, and the internal projectile wave function modified total reaction cross section of the Surrogate or Trojan Horse fragment $x$, \cite{CFH2017}, $\tilde{\sigma}^{(x + A)}_{R}$, tends, at higher energies, to the usual total reaction cross section $\sigma^{(x + A)}_{R}$. The resulting $b$ spectrum becomes the Serber cross section \cite{Serber1947}, $\sigma^{b}_{(Serber)} = K_{(Serber)} |\Phi_{a}(\textbf{k}_b)|^2 \sigma^{(x + A)}_{R}$. It is clear that the reaction cross section of the $x + A$ system is the sum of the direct cross section which is the sum of processes that proceed with short time delay and the compound nucleus cross section (incomplete fusion of $a$ or complete fusion of $x$) corresponding to the sum of statistical processes that proceed with long time delay, $\tilde{\sigma}^{(x + A)} = \tilde{\sigma}^{(x + A)}_{D} +\tilde{\sigma}^{(x + A)}_{CN}$. Each of these components is a sum of exclusive processes, $\tilde{\sigma}^{x + A}_{D} = \sum_{i} \tilde{\sigma}(x + A \rightarrow (y + B)_i)$, while $\tilde{\sigma}^{(x + A)}_{CN} = \sum_{i}K_{CN}\times[G_{(x + A)} T_{i}/\sum_{j}T_{j}]$. The Surrogate Method involves the extraction of the neutron capture cross section in a $(d,p)$ reaction. This amounts to use the INEB cross section for the observed proton in this reaction and extract from the data,
$\tilde{\sigma}^{(n + A)}_{R} = \sigma^{(p)}_{(INEB)}/K_{(INEB)}$. In view of the discussion given above, this cross section may contain a direct piece and this has to be calculated and subtracted from the reaction cross section to obtain the capture cross section. In the case of $n + ^{238}U$, the direct component can be estimated by performing a coupled channels calculation of inelastic excitation of the rotational states in this deformed target nucleus. Recent works \cite{Ducasse2015,Potel2015, Moro2015, Carlson2015}, were directed towards the justification of the SM cross section using the INEB theory of \cite{UT1981, IAV1985, HM1985, Austern1987}. For a recent review of this topic I direct the reader to \cite{Potel2017}.

The THM is contained in $\tilde{\sigma}^{(x + A)}_{D}$. The exclusive process of interest $A(x, y)B$, is one of the reactions that contributes to $\tilde{\sigma}^{(x + A)}_{D}$. Of course, in practice, several approximations are used to simplify the analysis, such as approximating the effect of the surrogate nature of $x$, namely its off shell nature, by adding its binding energy in $a$ to its laboratory energy, replacing $\tilde{\sigma}^{(x + A)}_{D}$ by $\sigma^{(x + A)}_{D}$, and accounting for the Fermi motion of $b$ through the density $|\phi(\textbf{k}_b)|^2$. 

Finally, the assumptions that the SM involves the compound nucleus formation and decay of the $n +A$ system, while the THM involves the direct reaction process $x + A \rightarrow y +B$ are only approximate as in general both processes may contain both compound and direct contributions. This is clearly manifest in the INEB theory, which supplies the internal primary projectile structure modified total reaction cross section of the $n + A$ or $x + A$ subsystem. 

I hope that in this note I have elucidated the similarities between the Surrogate Method and the Trojan Horse Method. Practitioners of  these methods would certainly benefit from the experience of each others and the community at large would be the winner.\\

{\it Acknowledgements.} 
I thank Jutta Escher and Aurora Tumino for very clarifying discussions on the SM and the THM. This work was partly supported by the Brazilian agencies, Funda\c c\~ao de Amparo \`a Pesquisa do Estado de
 S\~ao Paulo (FAPESP), the 
Conselho Nacional de Desenvolvimento Cient\'ifico e Tecnol\'ogico  (CNPq). I also acknowledge a Senior Visiting Professorship granted by the Coordena\c c\~ao de Aperfei\c coamento de Pessoal de N\'ivel Superior (CAPES), through the CAPES/ITA-PVS program.


\begin{thebibliography}{99}
\bibitem{SM0}Jutta E. Escher and Frank S. Dietrich, Phys. Rev. C \textbf{74}, 054601 (2006)
\bibitem{SM1} Jutta E. Escher, Jason T. Burke, Frank S. Dietrich, Nicholas D. Scielzo,
Ian J. Thompson, and Walid Younes, Rev. Mod. Phys., \textbf{84}, 353 (2012)
\bibitem{SM2}J. E. Escher, A. P. Tonchev, J. T. Burke, P. Bedrossian, R. J. Casperson, N. Cooper,, R.
O. Hughes, P. Humby, R. S. Ilieva,, S. Ota1,, N. Pietralla,, N. D. Scielzo, and V.Werner, EPJ Web of Conferences \textbf{122}, 12001 (2016)
\bibitem{Baur2003} S. Typel and G. Baur, Ann. Phys. (NY)  \textbf{305},  228 (2003)
\bibitem{THM1}C. Spitaleri et al., Phys. of Atomic Nuclei \textbf{74}, 1725 (2011) 
\bibitem{THM2} A. Tumino et al., Few Body Systems, \textbf{54}, 745 (2012) 
\bibitem{THM3} R. Tribble et al., Rep. Prog. Pays. \textbf{77}, 106901(2014) 
\bibitem{THM4} Jutta E. Escher and Frank S. Dietrich, Phys. Rev. C \textbf{81}, 024612 (2010) 
\bibitem{IAV1985}M. Ichimura, N. Austern, and C. M. Vincent, Phys. Rev. C \textbf{32}, 431 (1985)
\bibitem{UT1981}T. Udagawa and T. Tamura, Phys. Rev. C \textbf{24}, 1348 (1981)
\bibitem{HM1985}M. S. Hussein and K. W. McVoy, Nucl. Phys. A \textbf{445}, 124 (1985)
\bibitem{Austern1987}N. Austern, Y. Iseri, M. Kamimura, M. Kawai, G. Rawitscher,
and M. Yahiro, Phys. Rep. \textbf{154}, 125 (1987)
\bibitem{CFH2017} B.V. Carlson, T. Frederico, M.S. Hussein, Physics Letters B \textbf{767}, 53  (2017).
\bibitem{Serber1947} R. Serber, Phys. Rev. \textbf{12}, 1008 (1947) 
\bibitem{Ducasse2015} Q. Ducasse et al., Phys. Rev. C \textbf{94}, 024614 (2016)
\bibitem{Potel2015}G. Potel, F. M. Nunes, and I. J. Thompson, Phys. Rev. C \textbf{92}, 034611 (2015)
\bibitem{Moro2015} J.  Lei and A. M. Moro, Phys. Rev. C \textbf{92}, 044616 (2015);
 Phys. Rev. C \textbf{92}, 061602(R) (2015)
\bibitem{Carlson2015} B.V. Carlson, R. Capote, M. Sin, Few-Body Syst. {\textbf 57}, 307 (2016)
\bibitem{Potel2017} G. Potel et al., \textit{Toward a complete theory for predicting inclusive
deuteron breakup away from stability}, to be published (2017)
\end{thebibliography}
\end{document}